\documentclass[12pt]{article}
\usepackage{euscript,amsfonts}
\begin{document}
\title{\Large\bf{Whether conformal supersymmetry is broken by quantum $p$-branes with exotic supersymmetry?}}
\author{D.V.~Uvarov${}^a$ and A.A.~Zheltukhin${}^{a,b}$\\
{\normalsize ${}^a$ Kharkov Institute of Physics and Technology, 61108 
Kharkov, Ukraine}\\
{\normalsize ${}^{b}$ Institute of Theoretical Physics, University of 
Stockholm, Albanova,}\\
{\normalsize  SE-10691 Stockholm, Sweden}}
\date{}
\maketitle

\begin{abstract}

 Classical and quantum symmetries of super $p$-branes preserving exotic  $\frac{3}{4}$ fraction  of $N=1 D=4$ global supersymmetry are studied.
 Classical realization of the algebra of global and world-volume
symmetries is constructed and its quantum generalizations are analyzed.
Established is that the status of  the conformal supersymmetry  $OSp(1|8)$ as a proper quantum symmetry of brane depends both on the choice of its vacuum state and the  associated  ordering of $\hat{\cal Q}$ and $\hat{\cal P}$ operators.

\end{abstract}

\section{Introduction}

Studying  the symmetries of M-theory is one of the sharp problems and it is 
under  wide investigations which reveal new connections \cite{dufstel}, 
\cite{dufli}, \cite{hull}  between space-time symmetries and BPS vacuum 
states permitted by the extensions of the Poincare superalgebra by the tensorial  central charges (TCC) $Z_{m_1...m_p}$ \cite{DAF}, \cite{VHVP}.
Using TCC and  coordinates associated with them has opened new 
views for the construction of supersymmetric  generalizations of field and  
string/brane models  \cite{Zizzi}-\cite{caib}.
An interesting effect of the TCC inclusion is a possibility to regulate the 
fraction of spontaneously broken supersymmetry in the models based on the  
extended superalgebras \cite{BL}-\cite{gah2}.
On the other hand, the (super)space-times enlarged by the TCC coordinate 
addition provide well known framework for correct formulation of higher spin 
field theory \cite{Fronsdal}, \cite{Vasiliev} and its connection with the  
conformal ${\cal N}=4$ super Yang-Mills theory \cite{Bo}, \cite{Witten}.
These promising results stimulate interest in studying local and  global 
quantum symmetries of extended objects propagating in the superspaces enlarged 
 by  TCC coordinates.

A new exactly solvable  twistor-like model of (super)string and super  
$p$-brane  linear in the derivatives and preserving $\frac{{\sf M}-1}{{\sf 
M}}$ fraction of $N=1$ supersymmetry in the enlarged superspace, similarly 
to the superparticle model  \cite{BL}, was studied in \cite{ZU},\cite{ZU03}. 
  Because of the $OSp(1|2{\sf M})$ global symmetry of the model, its static 
$p$-brane solution was formulated in terms of symplectic supertwistors 
previously used while studying superparticle models \cite{Fer}, \cite{Shir}, 
\cite{BL} and forming a subspace of the $Sp(2{\sf M})$ invariant symplectic 
space \cite{Fronsdal}, \cite{Vasiliev}.
For the particular case of $D=11$ the model \cite{ZU} is invariant under the 
$OSp(1|64)$ symmetry and it was analyzed in \cite{Bandos02} and further 
generalized in \cite{BAPV}.
The  general static solutions of the  model \cite{ZU} describe  string/brane 
configurations which spontaneously break  $OSp(1|2{\sf M})$ global conformal 
supersymmetry.  The partially  spontaneously broken character of the 
$OSp(1|2{\sf M})$ symmetry was  also observed  in the Wess-Zumino like super 
$p$-brane models \cite{BeZ} nonlinear in derivatives and preserving 
$\frac{{\sf M}-1}{{\sf M}}$ fraction of supersymmetry. The models  
\cite{BeZ} generate the Dirichlet boundary conditions for the superstrings 
and superbranes as a consequence of the  variational principle.
These results unified with the observation concerning nonlinear realization of
$OSp(1|64)$ supersymmetry in $D=11$ supergravity \cite{West} support the 
tempting conjecture \cite{Vasiliev} that string/M-theory is a spontaneously 
broken phase of higher spin gauge theory.

The present paper  continues the investigation  of the Hamiltonian
and  quantum structure of the model of tensionless super $p-$branes
\cite{ZU} started in \cite{UZ03}. This model is characterized by
the primary and secondary constraints  which form  a mixture of
the first- and the second-class ones. These constraints  were
covariantly split in \cite{UZ03}, where the corresponding Dirac
brackets were constructed and the D.B. and operator realizations 
 of the orthosymplectic algebra in  $D=4 \, N=1$
enlarged superspace were obtained.

Here we consider an alternative approach to studying the  Hamiltonian 
structure of the model  which is based on  the conversion procedure 
\cite{Fad}-\cite{EM}. The  application of this procedure gives a possibility 
to transform all the primary constraints into the  first-class constraints 
that efficiently  simplifies investigations of the  global and gauge 
 symmetries of $p$-brane on the quantum level.
As a result,  we obtain the properly modified expressions for the
generator densities of $OSp(1|8)$ superalgebra on the classical
level  ensuring validity of the  global $OSp(1|8)$ supersymmetry
in the phase space extended by conversion variables. Then we
consider operator realizations of the  constraints, identified with
the  generators of world-volume gauge symmetry,  and generators of the
generalized conformal supersymmetry. The question whether  the classical global symmetry $OSp(1|8)$ of the exotic 
superstring and superbranes is  explicitly broken by quantization is investigated. 

We show that the status of $OSp(1|8)$ conformal supersymmetry as a proper quantum symmetry of exotic BPS brane depends both on the choice of its vacuum state and the corresponding physical ordering of the coordinate and momentum operators.

\section{Tensionless super $p$-branes with TCC coordinates: Lagrangian
formulation and primary constraints}

A simple model describing  tensionless strings and $p-$branes evolving in
the
symplectic superspace ${\cal M}^{susy}_{\sf M}$ was recently  proposed in
\cite{ZU}. For
${\sf M}=2^{[\frac{D}{2}]}$ $(D=2,3,4\ mod 8)$ the superspace ${\cal
M}^{susy}_{\sf M}$ includes  standard  $D$-dimensional Minkowski space-time
coordinates $x_{ab}=x^m(\gamma_mC^{-1})_{ab}$ accompanied by the
Majorana spinor $\theta_a$ $(a=1,2,...,2^{[\frac{D}{2}]})$ and the tensor
central charge coordinates $z_{ab}$ additively unified with the
space-time coordinates in the symmetric
spin-tensor $Y_{ab}$. The action of the model \cite{ZU}
\begin{equation}\label{1}
S_p=\frac12\int d\tau d^p\sigma\ \rho^\mu U^aW_{\mu ab}U^b
\end{equation}
includes the world-volume pullback
\begin{equation}\label{2}
W_{\mu ab}=\partial_\mu
Y_{ab}-2i(\partial_\mu\theta_a\theta_b+\partial_\mu\theta_b\theta_a)
\end{equation}
of the supersymmetric Cartan differential one-form $W_{ab}=W_{\mu
ab}d\xi^\mu$, where
$\partial_\mu\equiv\frac{\partial}{\partial\xi^\mu}$ and
$\xi^\mu=(\tau,\sigma^M)$, $M=1,2,...,p$ are world-volume
coordinates. The local auxiliary Majorana spinor $U^a(\tau,\sigma^M)$
parametrizes the generalized momentum $P^{ab}=\rho^{\tau}U^aU^b$ of the
tensionless $p-$brane and $\rho^\mu(\tau,\sigma^M)$ is the world-volume
vector density providing the reparametrization invariance of $S_p$. The
supersymmetric and world-volume  reparametrization invariant action
(\ref{1}) has $({\sf M}-1)$ $\kappa-$symmetries
\begin{equation}\label{3}
\delta_\kappa\theta_a=\kappa_a,
\quad\delta_\kappa
Y_{ab}=-2i(\theta_a\kappa_b+\theta_b\kappa_a),\quad
\delta_{\kappa} U^a=0,
\end{equation}
where the transformation parameter $\kappa_a(\xi)$ is restricted by only one
real
condition
\begin{equation}\label{4}
U^a\kappa_a=0,
\end{equation}
as it follows from the transformation rules of the primary variables
$\theta_a$, $U_a$ and $Y_{ab}=x_{ab}+z_{ab}$. As a result, the action
(\ref{1})  preserves $\frac{{\sf M}-1}{{\sf M}}$ fraction of the original
$N=1$
global supersymmetry.

$D=4$ $N=1$ supersymmetry algebra takes  the  form
\begin{equation}\label{5}
\{Q_a,Q_b\}=(\gamma^m C^{-1})_{ab}P_m+i(\gamma^{mn}C^{-1})_{ab}Z_{mn}
\end{equation}
including the TCC two-form $Z_{mn}$ with the charge conjugation matrix $C$
chosen to be imaginary in the Majorana representation.  The string/brane
model (\ref{1}) is formulated in generalized $(4+6)$-dimensional
space ${\cal M}_{4+6}$ extended by the Grassmannian Majorana bispinor
$\theta_a$.
In the Weyl basis real symmetric $4\times4$ matrix
$Y_{ab}$ acquires the form
\begin{equation}\label{6}
\begin{array}{c}
Y_{ab}=x_{ab}+z_{ab}=x_m(\gamma^m
C^{-1})_{ab}+z_{mn}(\gamma^{mn}C^{-1})_{ab},\\[0.2cm]
Y_a{}^b=Y_{ad}C^{db}=\left(\begin{array}{cc} z_\alpha{}^\beta&
x_{\alpha\dot\beta}\\ \tilde x^{\dot\alpha\beta}& \bar
z^{\dot\alpha}{}_{\dot\beta}\\ \end{array}\right),
\end{array}
\end{equation}
and  the auxiliary Majorana spinor $U^a(\tau,\sigma^M)$ is presented as
\begin{equation}\label{7}
U_a={u_\alpha\choose \bar u^{\dot\alpha}}
\end{equation}
with the charge conjugation matrix $C$
\begin{equation}\label{8}
C^{ab}=\left(\begin{array}{cc} \epsilon^{\alpha\beta}& 0\\ 0&
\epsilon_{\dot\alpha\dot\beta} \end{array}\right).
\end{equation}
Then the action (\ref{1}) acquires the form
\begin{equation}\label{9}
S_p=\frac12\int d\tau d^p\sigma\ \rho^\mu
\left(2u^\alpha\omega_{\mu\alpha\dot\alpha}\bar
u^{\dot\alpha}+u^\alpha\omega_{\mu\alpha\beta}u^\beta+\bar
u^{\dot\alpha}\bar\omega_{\mu\dot\alpha\dot\beta}\bar u^{\dot\beta}\right),
\end{equation}
where the supersymmetric one-forms $\omega_{\mu\alpha\dot\alpha}$ and
$\omega_{\mu\alpha\beta}$ are
\begin{equation}
%\label{10}
\begin{array}{c}
\omega_{\mu\alpha\dot\alpha}=\partial_\mu
x_{\alpha\dot\alpha}+2i(\partial_\mu\theta_\alpha\bar\theta_{\dot\alpha}+\partial_\mu\bar\theta_{\dot\alpha}\theta_\alpha),\\[0.2cm]
\omega_{\mu\alpha\beta}=-\partial_\mu
z_{\alpha\beta}-2i(\partial_\mu\theta_\alpha\theta_\beta+\partial_\mu\theta_\beta\theta_\alpha),\\[0.2cm]
\bar\omega_{\mu\dot\alpha\dot\beta}=-\partial_\mu\bar
z_{\dot\alpha\dot\beta}-2i(\partial_\mu\bar\theta_{\dot\alpha}\bar\theta_{\dot\beta}+\partial_\mu\bar\theta_{\dot\beta}\bar\theta_{\dot\alpha}).\\
\end{array}
\end{equation}

Following \cite{UZ03} let us introduce the momenta densities
${\cal P}^{\mathfrak M}(\tau,\sigma^M)$
\begin{equation}\label{11}
{\cal P}^{\mathfrak M}=\frac{\partial L}{\partial\dot{\cal
Q}_{\mathfrak M}}=(P^{\dot\alpha \alpha}, \pi^{\alpha\beta},
\bar\pi^{\dot\alpha\dot\beta}, \pi^\alpha, \bar\pi^{\dot\alpha}, P^\alpha_u,
\bar P^{\dot\alpha}_u, P^{(\rho)}_\mu)
\end{equation}
canonically conjugate to the coordinates ${\cal Q}_{\mathfrak
M}=(x_{\alpha\dot\alpha}, z_{\alpha\beta}, \bar
z_{\dot\alpha\dot\beta}, u_\alpha, \bar u_{\dot\alpha}, \rho^{\mu} )$
with respect to the Poisson brackets
\begin{equation}\label{12}
\{{\cal P}^{\mathfrak M}(\vec\sigma),{\cal Q}_{\mathfrak
N}(\vec\sigma')\}_{P.B.}=\delta^{\mathfrak M}_{\mathfrak
N}\delta^p(\vec\sigma-\vec\sigma ')
\end{equation}
with the periodic $\delta-$function $\delta^p(\vec\sigma-\vec\sigma')$, 
where $\vec\sigma=(\sigma^1,...,\sigma^p)$, for
the case of closed string/brane studied here.
The explicit form of P.B. (\ref{12}) is given by the expressions
\begin{equation}\label{17}
\begin{array}{c}
\{P^{\dot\alpha\alpha}(\vec\sigma),x_{\beta\dot\beta}(\vec\sigma^\prime)\}_{P.B.}=\delta^\alpha_\beta
\delta^{\dot\alpha}_{\dot\beta}\delta^p(\vec\sigma-\vec\sigma^\prime);
\\[0.2cm]
\{\pi^{\alpha\beta}(\vec\sigma),z_{\gamma\delta}(\vec\sigma^\prime)\}_{P.B.}=
\frac{1}{2}(\delta^\alpha_\gamma\delta^\beta_\delta +
\delta^{\alpha}_{\delta}
\delta^{\beta}_{\gamma})
\delta^p(\vec\sigma-\vec\sigma^\prime);
\\[0.2cm]
\{\bar\pi^{\dot\alpha\dot\beta}(\vec\sigma),\bar z_{\dot\gamma\dot\delta}
(\vec\sigma^\prime)\}_{P.B.}=\frac{1}{2}(\delta^{\dot\alpha}_{\dot\gamma}
\delta^{\dot\beta}_{\dot\delta}+
\delta^{\dot\alpha}_{\dot\delta}
\delta^{\dot\beta}_{\dot\gamma})
\delta^p(\vec\sigma-\vec\sigma^\prime);
\\[0.2cm]
\{\pi^\alpha(\vec\sigma),\theta_\beta(\vec\sigma^\prime)\}_{P.B.}=\delta^{\alpha}_{\beta}\delta^p(\vec\sigma-\vec\sigma^\prime),\quad
\{\bar\pi^{\dot\alpha}(\vec\sigma),\bar\theta_{\dot\beta}(\vec\sigma^\prime)\}_{P.B.}=
\delta^{\dot\alpha}_{\dot\beta}\delta^p(\vec\sigma-\vec\sigma^\prime);
\\[0.2cm]
\{P^{\alpha}_{u}(\vec\sigma),u_\beta(\vec\sigma^\prime)\}_{P.B.}=\delta^\alpha_\beta\delta^p(\vec\sigma-\vec\sigma^\prime),\quad
\{\bar P^{\dot\alpha}_{u}(\vec\sigma),\bar
u_{\dot\beta}(\vec\sigma^\prime)\}_{P.B.}=\delta^{\dot\alpha}_{\dot\beta}\delta^p(\vec\sigma-\vec\sigma^\prime);
\\[0.2cm]
\{P^{(\rho)}_\mu(\vec\sigma),\rho^\nu(\vec\sigma^\prime)\}_{P.B.}=\delta_\mu^\nu\delta^p(\vec\sigma-\vec\sigma^\prime).
\end{array}
\end{equation}

The primary constraints characterizing the action (\ref{9}) include
Grassmannian $\Psi$-constraints
\begin{eqnarray}\label{13}
\Psi^\alpha=\pi^\alpha-2i\bar\theta_{\dot\alpha}P^{\dot\alpha\alpha}-4i\pi^{\alpha\beta}\theta_\beta\approx0,\nonumber\\
\bar\Psi^{\dot\alpha}=-(\Psi^\alpha)^\ast=\bar\pi^{\dot\alpha}-2iP^{\dot\alpha\alpha}\theta_\alpha-4i\bar\pi^{\dot\alpha\dot\beta}\bar\theta_{\dot\beta}\approx0,
\end{eqnarray}
together with bosonic $\Phi$-constraints
\begin{equation}\label{14}\begin{array}{c}
\Phi^{\dot\alpha\alpha}=P^{\dot\alpha\alpha}-\rho^\tau u^\alpha
\bar u^{\dot\alpha}\approx0,\\[0.2cm]
\Phi^{\alpha\beta}=\pi^{\alpha\beta}+\frac12\rho^{\tau}u^\alpha
u^\beta\approx0,\\[0.2cm]
\bar\Phi^{\dot\alpha\dot\beta}=\bar\pi^{\dot\alpha\dot\beta}+\frac12\rho^{\tau}\bar
u^{\dot\alpha}
\bar u^{\dot\beta}\approx 0
\end{array}
\end{equation}
added by the constraints coming from the sector of  auxiliary world-volume
fields
$u_\alpha(\tau,\vec\sigma)$  and $\rho^\mu(\tau,\vec\sigma)$
\begin{equation}\label{15}
P^\alpha_u\approx0,\quad\bar P^{\dot\alpha}_u\approx0,
\end{equation}
\begin{equation}\label{16}
P^{(\rho)}_\tau\approx0,\quad P^{(\rho)}_M\approx0,
\end{equation}
where the world-volume index  $\mu=(\tau,M)$ was split into the
time-like and space-like $M=(1,2,...,p)$ subsets.

Using the definition of P.B. (\ref{17}) it was found in \cite{UZ03} that the
$\Psi$-constraints (\ref{13}) have zero Poisson
brackets with the primary bosonic constraints, but the P.B. of
$\Psi$-constraints among themselves are non-zero
\begin{eqnarray}\label{18}
\{\Psi^\alpha(\vec\sigma),\Psi^\beta(\vec\sigma{}')\}_{P.B.}=-8i\pi^{\alpha\beta}\delta^p(\vec\sigma-\vec\sigma{}')=-8i\left(\Phi^{\alpha\beta}-\frac12\rho^\tau
u^\alpha u^\beta\right)\delta^p(\vec\sigma-\vec\sigma{}'),\nonumber\\
\{\Psi^\alpha(\vec\sigma),\bar\Psi^{\dot\beta}(\vec\sigma{}')\}_{P.B.}=-4iP^{\dot\beta\alpha}\delta^p(\vec\sigma-\vec\sigma{}')=-4i\left(\Phi^{\dot\beta\alpha}+\rho^\tau
u^\alpha\bar u^{\dot\beta}\right)\delta^p(\vec\sigma-\vec\sigma{}'),
\nonumber\\
\{\bar\Psi^{\dot\alpha}(\vec\sigma),\bar\Psi^{\dot\beta}(\vec\sigma{}')\}_{P.B.}=-8i\bar\pi^{\dot\alpha\dot\beta}\delta^p(\vec\sigma-\vec\sigma{}')=
-8i\left(\bar\Phi^{\dot\alpha\dot\beta}-\frac12\rho^\tau
\bar u^{\dot\alpha}\bar
u^{\dot\beta}\right)\delta^p(\vec\sigma-\vec\sigma{}').
\end{eqnarray}
On the contrary the $\Phi$-constraints (\ref{14}) have zero Poisson
brackets among themselves and with the $\Psi-$constraints (\ref{13})
\begin{equation}\label{19}
\{\Phi(\vec\sigma),\Phi(\vec\sigma
')\}_{P.B.}=0,\quad\{\Phi(\vec\sigma),\Psi(\vec\sigma ')\}_{P.B.}=0,
\end{equation}
but do not commute with $P^\alpha_u$, $\bar P^{\dot\alpha}_u$
\begin{equation}\label{20}
\begin{array}{c}
\{P_{u}^\alpha(\vec\sigma), \Phi_{\beta\dot\gamma}(\vec\sigma')\}_{P.B.}=
-\rho^\tau
\delta^{\alpha}_{\beta}\bar u_{\dot\gamma}\delta^p(\vec\sigma-\vec\sigma'),
\\[0.2cm]
\{\bar P_{u}^{\dot\alpha}(\vec\sigma), 
\Phi_{\beta\dot\gamma}(\vec\sigma')\}_{P.B.}=
-\rho^\tau \delta^{\dot\alpha}_{\dot\gamma}u_\beta 
\delta^p(\vec\sigma-\vec\sigma'),
\\[0.2cm]
\{P_{u}^{\alpha}(\vec\sigma), 
\Phi_{\beta\gamma}(\vec\sigma')\}_{P.B.}=\frac12
\rho^\tau(\delta^{\alpha}_{\beta}u_\gamma+\delta^{\alpha}_{\gamma}u_\beta)\delta^p(\vec\sigma-\vec\sigma'),
\\[0.2cm]
\{\bar P_{u}^{\dot\alpha}(\vec\sigma), \bar
\Phi_{\dot\beta\dot\gamma}(\vec\sigma')\}_{P.B.}=\frac12\rho^\tau(\delta^{\dot\alpha}_{\dot\beta}\bar
u_{\dot\gamma}+
\delta^{\dot\alpha}_{\dot\gamma}
\bar u_{\dot\beta})
\delta^p(\vec\sigma-\vec\sigma')
\end{array}
\end{equation}
and $P^{(\rho)}_\tau$ (\ref{15})
\begin{equation}\label{21}
\begin{array}{c}
\{P^{(\rho)}_\tau(\vec\sigma), 
\Phi^{\beta\dot\gamma}(\vec\sigma')\}_{P.B.}=-u^\beta
\bar u^{\dot\gamma}\delta^p(\vec\sigma-\vec\sigma'),
\\[0.2cm]
\{P^{(\rho)}_\tau(\vec\sigma), \Phi^{\alpha\beta}(\vec\sigma')\}_{P.B.}=
\frac12 u^\alpha  u^\beta\delta^p(\vec\sigma-\vec\sigma'),
\\[0.2cm]
\{P^{(\rho)}_\tau(\vec\sigma), \bar
\Phi^{\dot\alpha\dot\beta}(\vec\sigma')\}_{P.B.}=
\frac12\bar u^{\dot\alpha}\bar 
u^{\dot\beta}\delta^p(\vec\sigma-\vec\sigma').
\end{array}
\end{equation}

It was shown in \cite{UZ03} that there are 3 first-class constraints among
$\Psi$-constraints and 6 first-class constraints among $\Phi$-constraints.
The momenta  $P^{(\rho)}_M\approx0$ also belong to the first class as it 
follows
from the observation that the conjugate coordinates $\rho^M$ do not enter
the expressions for the primary constraints.
Besides that there appear $(p+1)$
  extra first-class constraints $\Delta_W$ and  $L_{M}$
%as the sums of the primary constraints from
%different sectors. In particular,
corresponding to  the
world-volume Weyl and $\vec\sigma$-reparametrization gauge symmetries.
The Weyl symmetry affects only auxiliary fields $u_\alpha$
and $\rho^\mu$
\begin{equation}\label{WS}
\begin{array}{c}
\rho'^{\mu}=e^{-2\Lambda}\rho^\mu,\; u'_\alpha=e^\Lambda u_\alpha,
%\; v'_\alpha=e^{-\Lambda} v_\alpha,
\\
x'_{\alpha\dot\alpha}=x_{\alpha\dot\alpha},\
z'_{\alpha\beta}=z_{\alpha\beta},\ \theta'_\alpha= \theta_\alpha.
\end{array}
\end{equation}
It is generated by the first-class constraint
$\Delta_W$
\begin{equation}\label{WC}
\Delta_W\equiv(P^\alpha_u u_\alpha+\bar P^{\dot\alpha}_u\bar
u_{\dot\alpha})
%-(P^\alpha_v v_\alpha+\bar P^{\dot\alpha}_v\bar v_{\dot\alpha})
-2\rho^\mu P^{(\rho)}_\mu\approx0.
\end{equation}
The world-volume $\vec\sigma$-reparametrization constraints $L_{M}$ are 
given by
\begin{equation}\label{22}
\begin{array}{rl}
L_{M}=&P^{\dot\alpha\alpha}\partial_{M}x_{\alpha\dot\alpha}+\pi^{\alpha\beta}\partial_{M}z_{\alpha\beta}+\bar\pi^{\dot\alpha\dot\beta}\partial_{M}
\bar
z_{\dot\alpha\dot\beta}+\partial_{M}\theta_\alpha\pi^\alpha+\partial_{M}\bar\theta_{\dot\alpha}\bar\pi^{\dot\alpha}
\\[0.2cm]
+& P^{\alpha}_u\partial_M u_\alpha+\bar P^{\dot\alpha}_u
\partial_M\bar u_{\dot\alpha}- \partial_M P^{(\rho)}_{\tau}\rho^{\tau}
-\partial_M P^{(\rho)}_{N}\rho^{N}\approx0.
\end{array}
\end{equation}
They are secondary constraints  resulting from  the Dirac
selfconsistency procedure.

The Hamiltonian structure formed by the above defined constraints
after their covariant division into the first and second classes,
as well as the corresponding Dirac brackets (D.B.) were studied in
\cite{UZ03}. The procedure of the covariant constraint division
used in \cite{UZ03} involved additional auxiliary spinor field
$v_\alpha(\tau,\vec\sigma)$ forming together with
$u_\alpha(\tau,\vec\sigma)$ local spinorial basis attached to
string/brane world-volume.
As a result, the  additional first-class
constraints $P^{(u)}_v$ and $\bar P^{(u)}_v$
\begin{equation}\label{23}
P^{(u)}_v\equiv P^\alpha_v u_\alpha\approx0,\quad\bar P^{(u)}_v\equiv\bar
P^{\dot\alpha}_v\bar u_{\dot\alpha}\approx0
\end{equation}
arose corresponding to  the gauge shifts of $v_\alpha$  along $u_\alpha$
\begin{equation}\label{24}
\delta_\epsilon v_\alpha=\epsilon u_\alpha,\quad\delta_\epsilon\bar
v_{\dot\alpha}=\bar\epsilon\bar u_{\dot\alpha}.
\end{equation}
These shifts preserve the linear independence condition $ u^\alpha 
v_\alpha=1$
of the basis spinors  $ u^\alpha,\, v^\alpha$ and their D.B. turns out to be 
quadratic in the first class constraints
$\tilde T^{(\pm)},\, \Psi^{(u)}$ and its c.c.
\begin{equation}\label{25}
\begin{array}{rl}
\{P^{(u)}_v(\vec\sigma), \bar P^{(u)}_v(\vec\sigma')\}_{D.B.}
=&\frac{1}{4\rho^{\tau}}
[\tilde T^{(+)}(\bar P^{(u)}_v - P^{(u)}_v )+
\tilde T^{(-)}(\bar P^{(u)}_v + P^{(u)}_v ) \\[0.2cm]
+&\frac{1}{4}\Psi^{(u)}\bar\Psi^{(u})]\delta^p(\vec\sigma-\vec\sigma').
\end{array}
\end{equation}
This shows that the  D.B. algebra of the first-class constraints has the 
rank equal two and it gives rise to the higher powers of the ghosts in the 
BRST generator similarly to the (super) $p$-brane theory without TCC 
coordinates \cite{BZ}. By the  analogy the quadratic terms in the 
first-class constraints appear in the r.h.s. of D.B's  of the $OSp(1|8)$  
superalgebra.

Such a non-linear character of the both algebra of constraints and global 
$OSp(1|8)$ supersymmetry algebra complicates transition to  the quantum 
theory.
It gives a reason to apply the conversion
method  \cite{Fad}-\cite{EM} transforming  all the primary and secondary 
constraints in the first class.
This conversion method might turn out to be more efficient, similarly to the 
case of (super)particles \cite{ES2}-\cite{BLS}, and we  will utilize it  
here.

In the next Section we consider conversion of the primary and
secondary constraints (\ref{13})-(\ref{16}), (\ref{WC}), (\ref{22}) to the
first class and examine quantum realization of the symmetries of
the  model under question.

\section{Conversion of the primary and secondary constraints to the first
class}

To convert the constraints (\ref{15}) and (\ref{16}) to the first class we
introduce new variables
$\tilde u^\alpha=u^\alpha-q^\alpha$ and
$\tilde\rho^{\tau}=\rho^{\tau}-\varphi^{\tau}$, where $q^\alpha$ and
$\varphi^{\tau}$ are auxiliary conversion variables canonically conjugate to
momenta $P^\alpha_q$ and $P^{(\varphi)}_\tau$
\begin{equation}\label{28}
\begin{array}{c}
\{P^\alpha_q(\vec\sigma),q_\beta(\vec\sigma')\}_{P.B.}=\delta^\alpha_\beta\delta^p(\vec\sigma-\vec\sigma'),\quad\{
\bar P^{\dot\alpha}_q(\vec\sigma),
\bar
q_{\dot\beta}(\vec\sigma')\}_{P.B.}=\delta^{\dot\alpha}_{\dot\beta}\delta^p(\vec\sigma-\vec\sigma'),\\[0.2cm]
\{P^{(\varphi)}_{\tau}(\vec\sigma),\varphi^{\tau}(\vec\sigma')\}_{P.B.}=\delta^p(\vec\sigma-\vec\sigma').
\end{array}
\end{equation}
In terms of new  momenta the converted  constraints (\ref{15}) and
(\ref{16}) are presented as
\begin{eqnarray}\label{29}
\widetilde P^\alpha_u=P^\alpha_u+P^\alpha_q\approx0,\quad\bar{\widetilde
P}{}^{\dot\alpha}_u=\bar P^{\dot\alpha}_u+\bar P^{\dot\alpha}_q\approx0,
\end{eqnarray}
\begin{equation}\label{30}
\begin{array}{c}
\widetilde P^{(\rho)}_\tau=P^{(\rho)}_\tau+P^{(\varphi)}_\tau\approx0,\\
P^{(\rho)}_M\approx0
\end{array}
\end{equation}
after the correspondent modifications of the $\Phi-$
and  $\Psi$- sector constraints.
The converted  bosonic constraints
$\widetilde\Phi\equiv(\widetilde\Phi^{\dot\alpha\alpha},
\widetilde\Phi^{\alpha\beta},
\bar{\widetilde\Phi}{}^{\dot\alpha\dot\beta})$  originating from the
$\Phi$-constraints (\ref{14})
acquire the form
\begin{equation}\label{31}\begin{array}{c}
\widetilde\Phi^{\dot\alpha\alpha}=P^{\dot\alpha\alpha}-\tilde\rho^\tau\tilde
u^\alpha
\bar{\tilde u}{}^{\dot\alpha}\approx0,\\[0.2cm]
\widetilde\Phi^{\alpha\beta}=\pi^{\alpha\beta}+\frac12\tilde\rho^{\tau}\tilde
u^\alpha
\tilde u^\beta\approx0,\\[0.2cm]
\bar{\widetilde\Phi}{}^{\dot\alpha\dot\beta}=\bar\pi^{\dot\alpha\dot\beta}+\frac12\tilde\rho^{\tau}\bar{\tilde
u}{}^{\dot\alpha}
\bar{\tilde u}{}^{\dot\beta}\approx 0,
\end{array}
\end{equation}
and have  zero P.B. with the converted constraints  (\ref{29}), (\ref{30})
and among themselves.
Then the converted fermionic constraints
$\widetilde\Psi=(\widetilde\Psi^\alpha,
\bar{\widetilde\Psi}{}^{\dot\alpha})$ originating from the
$\Psi$-constraints
(\ref{13}) take  the form
\begin{equation}\label{32}\begin{array}{c}
\widetilde\Psi^\alpha=\pi^\alpha-2i\bar\theta_{\dot\alpha}P^{\dot\alpha\alpha}-4i\pi^{\alpha\beta}\theta_\beta+2(\tilde\rho^\tau)^{1/2}\tilde
u^\alpha f\approx0,\\
\bar{\widetilde\Psi}{}^{\dot\alpha}=-(\Psi^\alpha)^\ast=\bar\pi^{\dot\alpha}-2iP^{\dot\alpha\alpha}\theta_\alpha-4i\bar\pi^{\dot\alpha\dot\beta}\bar\theta_{\dot\beta}-2(\tilde\rho^\tau)^{1/2}\bar{\tilde
u}{}^{\dot\alpha} f\approx0,
\end{array}
\end{equation}
where $f^\ast=f$ is an  auxiliary Grassmannian  variable
characterized by the P.B.
\begin{equation}\label{33}
\{f(\vec\sigma),f(\vec\sigma')\}_{P.B.}=-i\delta^p(\vec\sigma-\vec\sigma').
\end{equation}
The addition of $f$  restores the forth $\kappa$-symmetry and
transforms  the $\widetilde\Psi$-constraints to the  first class
leading to non-zero  P.B. only among themselves
\begin{equation}\label{34}
\{\widetilde\Psi^\alpha(\vec\sigma),\widetilde\Psi^\beta(\vec\sigma{}')\}_{P.B.}=-8i\widetilde\Phi^{\alpha\beta}\delta^p(\vec\sigma-\vec\sigma{}'),
\end{equation}
\begin{equation}\label{35}
\{\widetilde\Psi^\alpha(\vec\sigma),\bar{\widetilde\Psi}{}^{\dot\beta}(\vec\sigma{}')\}_{P.B.}=-4i\widetilde\Phi^{\dot\beta\alpha}\delta^p(\vec\sigma-\vec\sigma{}').
\end{equation}

The  Weyl symmetry (\ref{WC}) and world-volume 
$\vec\sigma-$reparametrization
constraints in the extended  phase space  are given by
\begin{equation}\label{36}
\widetilde\Delta_W\equiv(\tilde{\tilde P}{}^\alpha_u\tilde
u_\alpha+\bar{\tilde{\tilde P}}{}^{\dot\alpha}_u
\bar{\tilde u}_{\dot\alpha})-2\tilde\rho^\tau\tilde{\tilde
P}{}^{(\rho)}_\tau-2\rho^M P^{(\rho)}_M\approx0,
\end{equation}
where
\begin{equation}\label{37}
\tilde{\tilde P}{}^\alpha_u=\frac12(P^\alpha_u-P^\alpha_q), \quad \quad
\tilde{\tilde
P}{}^{(\rho)}_\tau=\frac12(P^{(\rho)}_\tau-P^{(\varphi)}_\tau),
\end{equation}
and respectively
\begin{equation}\label{38}
\begin{array}{rl}
\widetilde
L_{M}=&P^{\dot\alpha\alpha}\partial_{M}x_{\alpha\dot\alpha}+\pi^{\alpha\beta}\partial_{M}z_{\alpha\beta}+\bar\pi^{\dot\alpha\dot\beta}\partial_{M}
\bar
z_{\dot\alpha\dot\beta}+\partial_{M}\theta_\alpha\pi^\alpha+\partial_{M}\bar\theta_{\dot\alpha}\bar\pi^{\dot\alpha}
\\[0.2cm]
+&(\tilde{\tilde P}{}^\alpha_u\partial_M\tilde u_\alpha+\bar{\tilde{\tilde
P}}{}^{\dot\alpha}_u\partial_M\bar{\tilde u}_{\dot\alpha})\\[0.2cm]
-&\partial_M
\tilde{\tilde P}{}^{(\rho)}_{\tau}\tilde\rho^{\tau}-\partial_M
P^{(\rho)}_{N}\rho^{N}-\frac{i}{2}f\partial_M f\approx0.
\end{array}
\end{equation}
These converted constraints satisfy the following non zero P.B.
relations
\begin{equation}\label{39}
\{\widetilde\Delta_W(\vec\sigma),P^{(\rho)}_M(\vec\sigma')\}_{P.B.}=2P^{(\rho)}_M\delta^p(\vec\sigma-\vec\sigma'),
\end{equation}
\begin{equation}\label{40}
\{\widetilde L_M(\vec\sigma),\widetilde
P^{\alpha}_u(\vec\sigma')\}_{P.B.}=0,
\end{equation}
\begin{equation}\label{41}
\{\widetilde L_M(\vec\sigma),\widetilde
P^{(\rho)}_\tau(\vec\sigma')\}_{P.B.}=0,
\end{equation}
\begin{equation}\label{42}
\{\widetilde L_M(\vec\sigma),
P^{(\rho)}_N(\vec\sigma')\}_{P.B.}=\partial_M\tilde
P^{(\rho)}_N\delta^p(\vec\sigma-\vec\sigma'),
\end{equation}
\begin{equation}\label{43}
\begin{array}{rl}
\{\widetilde L_M(\vec\sigma),\widetilde
L_N(\vec\sigma')\}_{P.B.}=&(\widetilde
L_M(\vec\sigma')\partial_{N'}-\widetilde
L_N(\vec\sigma)\partial_M)\delta^p(\vec\sigma-\vec\sigma')
%\\[0.2cm]=&-(\widetilde L_M(\vec\sigma)\partial_N+\widetilde
%L_N(\vec\sigma)\partial_M+\partial_N\widetilde
%L_M(\vec\sigma))\delta^p(\vec\sigma-\vec\sigma')
,
\end{array}
\end{equation}
\begin{equation}\label{44}
\{\widetilde L_M(\vec\sigma),[\mbox{other
constraints}](\vec\sigma')\}_{P.B.}=-[\mbox{other
constraints}](\vec\sigma)\partial_M\delta^p(\vec\sigma-\vec\sigma')
\end{equation}
and the complex conjugate ones. Other P.B's. are equal to zero in the strong 
  sense.

One can easily verify that there was added such the number of conversion
variables
so that the number of physical degrees of freedom in the enlarged phase
space
is equal to
\begin{equation}\label{45}
\begin{array}{rl}
n^{\Phi}_{\mbox{{\footnotesize phys}}}=& 2(15_b+4_f)_{\mbox{{\footnotesize
original}}}+2(5_b+{\textstyle\frac12} 1_f)_{\mbox{{\footnotesize
conversion}}}-2((16+p)_b+4_f)_{\mbox{{\footnotesize $1^{\underline st}$ cl.
constr.}}}\\[0.2cm]
=& 2(4-p)_b+1_f,
\end{array}
\end{equation}
and it  coincides with the  number of physical degrees of freedom in the
original model
\begin{equation}\label{46}
\begin{array}{rl}
n_{\mbox{{\footnotesize phys}}}=& 2(15_b+4_f)_{\mbox{{\footnotesize
original}}}-2((7+p)_b+3_f)_{\mbox{{\footnotesize $1^{\underline st}$ cl.
constr.}}}-(8_b+1_f)_{\mbox{\footnotesize $2^{\underline nd}$ cl.
constr.}}\\[0.2cm]
=& 2(4-p)_b+1_f.
\end{array}
\end{equation}
Note that  we have not taken into account the variables $\rho_{M}$
which are involved  in the original and  extended phase spaces,
because of the first-class constraints $P^{(\rho)}_M$ (\ref{30})
presence in both of these phase spaces. Thus, we have solved the
conversion problem and found  the first-class constraints  in the
extended phase space. The next step is to find
a realization of the $OSp(1|8)$ global supersymmetry generators in the
extended phase space.

\section{Generators of the $OSp(1|8)$ global supersymmetry in the extended
phase space}

Here we shall construct the  $OSp(1|8)$ generators in the extended phase
space
based on their representation built  in \cite {UZ03}.
The key principle to find these extended generators is the requirement for
them
to form a closed P.B. algebra with the above considered converted
first-class constraints. This requirement together with the requirement to
generate $OSp(1|8)$ superalgebra in the extended phase space will uniquely
restore the form of these generators.

We find  that the generalized translation generators together the  generator
densities $Q^\alpha$ and $Q^{\dot\alpha}$ of the $N=1$ global supersymmetry
remain unchanged upon transition to extended phase space
\begin{eqnarray}\label{47}
Q^{\alpha}(\tau,\vec\sigma)=\pi^\alpha +2i\bar\theta_{\dot\alpha}
P^{\dot\alpha\alpha}+
4i\pi^{\alpha\beta}\theta_{\beta}, \nonumber  \\
\bar Q^
{\dot\alpha}(\tau,\vec\sigma)=\bar\pi^{\dot\alpha}
+2iP^{\dot\alpha\alpha}\theta_{\alpha}+
4i\bar\pi^{\dot\alpha\dot\beta}\bar\theta_{\dot\beta}
\end{eqnarray}
and their P.B's. have the standard form
\begin{eqnarray}\label{48}
\{ Q^\alpha(\vec\sigma), \bar Q^{\dot\alpha}(\vec\sigma')\}_{P.B.}
=4iP^{\dot\alpha\alpha}\delta^p(\vec\sigma-\vec\sigma'), \nonumber  \\
\{ Q^{\alpha}(\vec\sigma), Q^{\beta}(\vec\sigma')\}_{P.B.}=8i
\pi^{\alpha\beta}\delta^p(\vec\sigma-\vec\sigma').
\end{eqnarray}
To build the extended "square roots" $\widetilde S_\gamma$ and
$\bar{\widetilde S}_{\dot\gamma}$ of the generalized conformal boost
densities
%$K_{\gamma\dot\gamma}$, $K_{\gamma\lambda}$
we consider the  P.B. relations
$\{\widetilde\Psi^\alpha(\vec\sigma),S_\gamma(\vec\sigma')\}_{P.B.}$ and 
find
that the additional term $\frac{2}{(\tilde\rho^\tau)^{1/2}}\tilde{\tilde
P}_{u\gamma}f$ has to be added to  $S_\gamma$ to close the P.B.
\begin{equation}\label{49}
\{\widetilde\Psi^\alpha(\vec\sigma),\widetilde S_\gamma(\sigma')\}_{P.B.}=
4i\delta^{\alpha}_{\gamma}[\widetilde\Psi^\beta\theta_{\beta} +
\bar{\widetilde\Psi}{}^{\dot\beta}
\bar\theta_{\dot\beta}]\delta^p(\vec\sigma-\vec\sigma'),
\end{equation}
where  $\widetilde S_\gamma$  and its complex conjugate are given by
\begin{equation}\label{50}
\begin{array}{rl}
\widetilde S_{\gamma}(\tau,\vec\sigma)=z_{\gamma\delta}Q^\delta +
x_{\gamma\dot\delta}\bar
Q^{\dot\delta}-2i\theta_\gamma(\theta_\delta\pi^\delta+\bar\theta_{\dot\delta}\bar\pi^{\dot\delta})+4i(\tilde
u^\delta\theta_\delta-\bar{\tilde
u}{}^{\dot\delta}\bar\theta_{\dot\delta})\tilde{\tilde
P}_{u\gamma}+\frac{2}{(\tilde\rho^\tau)^{1/2}}\tilde{\tilde P}_{u\gamma}f,
\\
\bar{\widetilde S}_{\dot\gamma}(\tau,\vec\sigma)=\bar
z_{\dot\gamma\dot\delta}\bar
Q^{\dot\delta}+x_{\delta\dot\gamma}Q^{\delta}-2i\bar\theta_{\dot\gamma}(\theta_\delta\pi^\delta+\bar\theta_{\dot\delta}\bar\pi^{\dot\delta})-4i(\tilde
u^\delta\theta_\delta-\bar{\tilde
u}{}^{\dot\delta}\bar\theta_{\dot\delta})\bar{\tilde{\tilde
P}}{}_{u\dot\gamma}-\frac{2}{(\tilde\rho^\tau)^{1/2}}\bar{\tilde{\tilde
P}}{}_{u\dot\gamma}f.
\end{array}
\end{equation}

The  representations  (\ref{50}) can be used to find the generalized
conformal boost densities $\widetilde K_{\gamma\lambda}$ and $\widetilde
K_{\gamma\dot\gamma}$
from  the known P.B. relations of $OSp(1|8)$ superalgebra
\begin{equation}\label{51}
\{\widetilde S_\gamma(\vec\sigma),\widetilde
S_\lambda(\vec\sigma')\}_{P.B.}=4i\widetilde
K_{\gamma\lambda}\delta^p(\vec\sigma-\vec\sigma'),
\quad
\{\widetilde S_\gamma(\vec\sigma),\bar{\widetilde
S}_{\dot\gamma}(\vec\sigma')\}_{P.B.}
=4i\widetilde K_{\gamma\dot\gamma}\delta^p(\vec\sigma-\vec\sigma'),
\end{equation}
resulting in the explicit form for  the $K-$generators
\begin{equation}\label{52}
\begin{array}{rl}
\widetilde K_{\gamma\lambda}(\tau,\vec\sigma)=&
2z_{\gamma\beta}z_{\lambda\delta}\pi^{\beta\delta}+2x_{\gamma\dot\beta}x_{\lambda\dot\delta}\bar\pi^{\dot\beta\dot\delta}+z_{\gamma\beta}x_{\lambda\dot\delta}P^{\dot\delta\beta}+x_{\gamma\dot\beta}z_{\lambda\delta}
P^{\dot\beta\delta}\\[0.2cm]
+&
\theta_\lambda(z_{\gamma\delta}\pi^{\delta}+x_{\gamma\dot\delta}\bar\pi^{\dot\delta})+\theta_{\gamma}(z_{\lambda\delta}\pi^{\delta}+x_{\lambda\dot\beta}\bar\pi^{\dot\delta})\\[0.2cm]
+& (\tilde u^{\delta} z_{\delta\lambda}-\bar{\tilde u}{}^{\dot\delta}
x_{\lambda\dot\delta})\tilde{\tilde P}_{
u\gamma}+(\tilde u^\delta z_{\delta\gamma}-\bar{\tilde
u}{}^{\dot\delta}x_{\gamma\dot\delta})\tilde{\tilde P}_{u\lambda}\\[0.2cm]
-& 2i(\tilde u^\delta\theta_\delta-\bar{\tilde
u}{}^{\dot\delta}\bar\theta_{\dot\delta})(\theta_{\lambda}\tilde{\tilde
P}_{u\gamma}+\theta_\gamma\tilde{\tilde P}_{u\lambda})\\
+&\frac{2}{(\tilde\rho^\tau)^{1/2}}(\theta_{\lambda}\tilde{\tilde
P}_{u\gamma}+\theta_\gamma\tilde{\tilde
P}_{u\lambda})f-\frac{1}{\tilde\rho^\tau}\tilde{\tilde
P}_{u\gamma}\tilde{\tilde P}_{u\lambda},\\
\end{array}
\end{equation}
\begin{equation}\label{53}
\begin{array}{rl}
\widetilde K_{\gamma\dot\gamma}(\tau,\vec\sigma)=& z_{\gamma\delta}\bar
z_{\dot\gamma\dot\delta}P^{\dot\delta\delta}+x_{\gamma\dot\delta}x_{\delta\dot\gamma}P^{\dot\delta\delta}+2(z_{\gamma\delta}x_{\lambda\dot\gamma}\pi^{\delta\lambda}+x_{\gamma\dot\delta}\bar
z_{\dot\gamma\dot\lambda}\bar\pi^{\dot\delta\dot\lambda})\\[0.2cm]
+& \theta_\gamma(\bar
z_{\dot\gamma\dot\delta}\bar\pi^{\dot\delta}+x_{\delta\dot\gamma}\pi^{\delta})+\bar\theta_{\dot\gamma}(z_{\gamma\delta}\pi^{\delta}+x_{\gamma\dot\delta}\bar\pi^{\dot\delta})\\[0.2cm]
+& (\tilde u^{\delta} x_{\delta\dot\gamma}-\bar{\tilde u}{}^{\dot\delta}\bar
z_{\dot\delta\dot\gamma})\tilde{\tilde
P}_{u\gamma}+(x_{\gamma\dot\delta}\bar{\tilde
u}{}^{\dot\delta}-z_{\gamma\delta}\tilde u^\delta)\bar{\tilde{\tilde
P}}_{u\dot\gamma}\\[0.2cm]
-& 2i(\tilde u^\delta\theta_\delta-\bar{\tilde
u}{}^{\dot\delta}\bar\theta_{\dot\delta})(\bar\theta_{\dot\gamma}\tilde{\tilde
P}_{u\gamma}-\theta_\gamma \bar{\tilde{\tilde P}}_{u\dot\gamma})\\
+&\frac{2}{(\tilde\rho^\tau)^{1/2}}(\bar\theta_{\dot\gamma}\tilde{\tilde
P}_{u\gamma}-\theta_\gamma \bar{\tilde{\tilde
P}}_{u\dot\gamma})f+\frac{1}{\tilde\rho^\tau}\tilde{\tilde
P}_{u\gamma}\bar{\tilde{\tilde P}}_{u\dot\gamma}.\\
\end{array}
\end{equation}

The correctness of  the representations  (\ref{50}) is verified by the 
reproducibility of the known P.B. relations
\begin{equation}\label{54}
\{Q_\alpha(\vec\sigma),{\widetilde
K}_{\beta\gamma}(\vec\sigma')\}_{P.B.}=(\varepsilon_{\alpha\beta}
\widetilde S_{\gamma}+\varepsilon_{\alpha\gamma}
\widetilde S_\beta)\delta^p(\vec\sigma-\vec\sigma'),
\
\{Q_\alpha(\vec\sigma),{\widetilde
K}_{\beta\dot\gamma}(\vec\sigma')\}_{P.B.}=\varepsilon_{\alpha\beta}\bar
{\widetilde S}_{\dot\gamma}\delta^p(\vec\sigma-\vec\sigma').
\end{equation}

Remaining 16 generator densities of the $Sp(8)$ algebra $\widetilde
L^{\alpha}{}_{\beta}$, $\widetilde L^{\alpha}{}_{\dot\beta}$ that include 
the
Lorentz symmetry generator densities are given by the expressions
\begin{equation}\label{55}
\begin{array}{c}
\widetilde
L^{\alpha}{}_{\beta}(\tau,\vec\sigma)=P^{\dot\beta\alpha}x_{\beta\dot\beta}+
2\pi^{\alpha\gamma}z_{\gamma\beta} - \pi^\alpha\theta_\beta +\tilde
u^\alpha\tilde{\tilde P}_{u\beta},\\
\widetilde L^\alpha{}_{\dot\beta}(\tau,\vec\sigma)=
2\pi^{\alpha\gamma}x_{\gamma\dot\beta} +
P^{\dot\gamma\alpha}\bar z_{\dot\beta\dot\gamma}
- \pi^\alpha\bar\theta_{\dot\beta} -\tilde u^\alpha \bar{\tilde{\tilde
P}}_{u\dot\beta}
\end{array}
\end{equation}
and their complex conjugate,
which follow from the P.B. relations of the $OSp(1|8)$ superalgebra
\begin{eqnarray}\label{56}
\{Q_\alpha(\vec\sigma),\widetilde S_\beta(\vec\sigma')\}_{P.B.}=4i\widetilde
L_{\alpha\beta}\delta^p(\vec\sigma-\vec\sigma'),\quad\{Q_\alpha(\vec\sigma),\bar
{\widetilde S}_{\dot\beta}(\vec\sigma')\}_{P.B.}=4i\widetilde
L_{\alpha\dot\beta}\delta^p(\vec\sigma-\vec\sigma'),
\end{eqnarray}
where  the representations (\ref{47}) and (\ref{50}) have been used.

The correctness of the representations (\ref{55}) is verified by the 
reproducibility  of well known  P.B's. for the $L$-densities
\begin{equation}\label{57}
\begin{array}{c}
\{\widetilde L_{\alpha\beta}(\vec\sigma), \widetilde
L_{\gamma\delta}(\vec\sigma')\}_{P.B.}=(\varepsilon_{\beta\gamma}
\widetilde L_{\alpha\delta}+\varepsilon_{\alpha\delta}\widetilde
L_{\gamma\beta})\delta^p(\vec\sigma-\vec\sigma'),
\\[0.2cm]
\{\widetilde L_{\alpha\beta}(\vec\sigma), \widetilde
L_{\dot\gamma\delta}(\vec\sigma')\}_{P.B.}=\varepsilon_{\alpha\delta}
\widetilde L_{\dot\gamma\beta}\delta^p(\vec\sigma-\vec\sigma'),
\quad
\{\widetilde L_{\alpha\beta}(\vec\sigma),
\widetilde
L_{\gamma\dot\delta}(\vec\sigma')\}_{P.B.}=\varepsilon_{\beta\gamma}
\widetilde L_{\alpha\dot\delta}\delta^p(\vec\sigma-\vec\sigma'),
\\[0.2cm]
\{\widetilde L_{\alpha\dot\beta}(\vec\sigma),
\widetilde
L_{\dot\gamma\delta}(\vec\sigma')\}_{P.B.}=(\varepsilon_{\dot\beta\dot\gamma}
\widetilde L_{\alpha\delta}+\varepsilon_{\alpha\delta}\bar
{\widetilde L}_{\dot\gamma\dot\beta})\delta^p(\vec\sigma-\vec\sigma')
\end{array}
\end{equation}
added  by
their P.B's. with  the  densities of supercharges $Q$ and  $\widetilde S$
\begin{equation}\label{58}
\begin{array}{c}
\{Q_\alpha(\vec\sigma),\widetilde
L_{\beta\gamma}(\vec\sigma')\}_{P.B.}=\varepsilon_{\alpha\gamma}
Q_{\beta}\delta^p(\vec\sigma-\vec\sigma'),\quad\{Q_\alpha(\vec\sigma),\widetilde
L^{\dot\beta}{}_{\gamma}(\vec\sigma')\}_{P.B.}=-\varepsilon_{\alpha\gamma}\bar
Q^{\dot\beta}\delta^p(\vec\sigma-\vec\sigma'),
\\[0.2cm]
\{\widetilde S_\alpha(\vec\sigma),\widetilde
L_{\beta\gamma}(\vec\sigma')\}_{P.B.}=\varepsilon_{\alpha\beta}\widetilde
S_{\gamma}\delta^p(\vec\sigma-\vec\sigma'),\quad\{\widetilde
S_\alpha(\vec\sigma),\widetilde
L_{\beta}{}^{\dot\gamma}(\vec\sigma')\}_{P.B.}=
\varepsilon_{\alpha\beta}\bar{\widetilde
S}^{\dot\gamma}\delta^p(\vec\sigma-\vec\sigma')
\end{array}
\end{equation}
and with the conformal boost densities $\widetilde K$
\begin{equation}\label{59}
\begin{array}{c}
\{{\widetilde L}_{\alpha\beta}(\vec\sigma),
\widetilde
K_{\gamma\delta}(\vec\sigma')\}_{P.B.}=(\varepsilon_{\alpha\gamma}\widetilde
K_{\beta\delta}+\varepsilon_{\alpha\delta}
\widetilde K_{\beta\gamma})\delta^p(\vec\sigma-\vec\sigma'),
\\[0.2cm]
\{L_{\alpha\beta}(\vec\sigma),
\widetilde
K_{\gamma\dot\delta}(\vec\sigma')\}_{P.B.}=\varepsilon_{\alpha\gamma}\widetilde
K_{\beta\dot\delta}\delta^p(\vec\sigma-\vec\sigma'),
\\[0.2cm]
\{\widetilde L_{\alpha\dot\beta}(\vec\sigma),
\widetilde
K_{\gamma\delta}(\vec\sigma')\}_{P.B.}=(\varepsilon_{\alpha\gamma}\widetilde
K_{\delta\dot\beta}+\varepsilon_{\alpha\delta}
\widetilde K_{\gamma\dot\beta})\delta^p(\vec\sigma-\vec\sigma'),
\\[0.2cm]
\{\widetilde L_{\alpha\dot\beta}(\vec\sigma),
\widetilde
K_{\gamma\dot\delta}(\vec\sigma')\}_{P.B.}=\varepsilon_{\alpha\gamma}\bar
{\widetilde K}_{\dot\beta\dot\delta}\delta^p(\vec\sigma-\vec\sigma'),
\end{array}
\end{equation}
where we adduced only nonzero Poisson brackets.

The remaining nonzero P.B's. of the  considered orthosymplectic superalgebra
include the
generator densities  $P^{\beta\dot\gamma}$, $\pi^{\beta\gamma}$  of
the generalized translations
\begin{equation}\label{60}
\begin{array}{c}
\{\widetilde
S_\alpha(\vec\sigma),\pi_\beta{}^\gamma(\vec\sigma')\}_{P.B.}=-\frac12(\delta_\alpha^\gamma
Q_\beta-\varepsilon_{\alpha\beta}Q^\gamma)\delta^p(\vec\sigma-\vec\sigma'),
\\[0.2cm]
\{\widetilde
S_\alpha(\vec\sigma),P_{\beta\dot\gamma}(\vec\sigma')\}_{P.B.}=-\varepsilon_{\alpha\beta}\bar
Q_{\dot\gamma}\delta^p(\vec\sigma-\vec\sigma');
\\[0.2cm]
\{\pi_\alpha{}^\beta(\vec\sigma),\widetilde
L_{\gamma\delta}(\vec\sigma')\}_{P.B.}=(\delta^\beta_\delta\pi_{\alpha\gamma}+\varepsilon_{\alpha\delta}\pi_{\gamma}{}^{\beta})\delta^p(\vec\sigma-\vec\sigma'),
\\[0.2cm]
\{\pi_\alpha{}^\beta(\vec\sigma),\widetilde
L_{\dot\gamma}{}_\delta(\vec\sigma')\}_{P.B.}=\frac12(\delta^\beta_\delta
P_{\alpha\dot\gamma}-\varepsilon_{\alpha\delta}P^{\beta}{}_{\dot\gamma})\delta^p(\vec\sigma-\vec\sigma'),
\\[0.2cm]
\{P_{\alpha\dot\beta}(\vec\sigma),{\widetilde
L}_{\gamma\delta}(\vec\sigma')\}_{P.B.}=\varepsilon_{\alpha\delta}P_{\gamma\dot\beta}\delta^p(\vec\sigma-\vec\sigma'),\
\{P_{\alpha\dot\beta}(\vec\sigma),{\widetilde
L}^{\dot\gamma}{}_{\delta}(\vec\sigma')\}_{P.B.}=2\varepsilon_{\alpha\delta}\bar\pi^{\dot\gamma}{}_{\dot\beta}\delta^p(\vec\sigma-\vec\sigma'),
\\[0.2cm]
\{\pi_\alpha{}^\beta(\vec\sigma),\widetilde
K_{\gamma\delta}(\vec\sigma')\}_{P.B.}=\frac12(\delta^\beta_\gamma
\widetilde L_{\alpha\delta}+\varepsilon_{\alpha\gamma}{\widetilde
L}^{\beta}{}_{\delta}+\delta_\beta^\delta
\widetilde L_{\alpha\gamma}+\varepsilon_{\alpha\delta} {\widetilde
L}^{\beta}{}_{\gamma})\delta^p(\vec\sigma-\vec\sigma'),
\\[0.2cm]
\{\pi_\alpha{}^\beta(\vec\sigma),
{\widetilde
K}_{\gamma\dot\delta}(\vec\sigma')\}_{P.B.}=-\frac12(\delta^\beta_\gamma
{\widetilde L}_{\alpha\dot\delta}-\varepsilon_{\alpha\gamma}{\widetilde
L}^{\beta}{}_{\dot\delta})\delta^p(\vec\sigma-\vec\sigma'),
\\[0.2cm]
\{P_{\alpha\dot\beta}(\vec\sigma), {\widetilde
K}_{\gamma\delta}(\vec\sigma')\}_{P.B.}=(\varepsilon_{\alpha\delta}
{\widetilde L}_{\dot\beta\gamma}+\varepsilon_{\alpha\gamma}{\widetilde
L}_{\dot\beta\delta})\delta^p(\vec\sigma-\vec\sigma'),\\[0.2cm]
\{P_{\alpha\dot\beta}(\vec\sigma),
\widetilde
K_{\gamma\dot\delta}(\vec\sigma')\}_{P.B.}=(\varepsilon_{\alpha\gamma}\bar
{\widetilde
L}_{\dot\beta\dot\delta}+\varepsilon_{\dot\beta\dot\delta}{\widetilde
L}_{\alpha\gamma})\delta^p(\vec\sigma-\vec\sigma').
\end{array}
\end{equation}
These expressions should be complemented by their complex conjugate.

So, one can conclude that  the  above considered P.B's.  correctly
reproduce the well known P.B. commutation relations of the
$OSp(1|8)$ superalgebra and, therefore, the problem of realization
of this superalgebra in the extended phase space of the model is
solved.

The next  step  is to find the  P.B. commutation relations between
the generators of  the $OSp(1|8)$ superalgebra and the generators
of the gauge symmetry of the model presented by the  first-class
constraints constructed in the Section 3.

\section{The P.B. superalgebra of global and gauge symmetries in the
extended phase space}

Taking into account that the $OSp(1|8)$ superalgebra generators
are identified with the physical observables their P.B. relations
with the converted first-class constraints have to be zero in the
weak sense.  Fulfilment of this condition would prove the gauge
invariant character of the  considered super $p$-brane model. In
this  section  we show  that this  condition is actually satisfied
on the classical level of the P.B. superalgebra.

At first,  we find that the
$\widetilde \Phi$-constraints have the following nonzero Poisson brackets
with
the $\widetilde L$ generator densities of $OSp(1|8)$
\begin{equation}\label{L}
\begin{array}{c}
\{\widetilde\Phi^{\alpha\beta}(\vec\sigma),
  {\widetilde
L}_{\gamma\delta}(\vec\sigma')\}_{P.B.}=(\delta^{\alpha}_{\delta}\widetilde\Phi_{\gamma}{}^{\beta}+\delta^{\beta}_{\delta}\widetilde\Phi_{\gamma}{}^{\alpha})\delta^p(\vec\sigma-\vec\sigma'),\\[0.2cm]
\{\widetilde\Phi^{\alpha\beta}(\vec\sigma),
L_{\dot\gamma\delta}(\vec\sigma')\}_{P.B.}=\frac12(\delta^{\alpha}_{\delta}\widetilde\Phi^{\beta}{}_{\dot\gamma}+\delta^{\beta}_{\delta}\widetilde\Phi^{\alpha}{}_{\dot\gamma})\delta^p(\vec\sigma-\vec\sigma');\\[0.2cm]
\{\widetilde\Phi^{\dot\alpha\alpha}(\vec\sigma),
L_{\gamma\delta}(\vec\sigma')\}_{P.B.}=\delta^{\alpha}_{\delta}\widetilde\Phi_{\gamma}{}^{\dot\alpha}\delta^p(\vec\sigma-\vec\sigma'),\\[0.2cm]
\{\widetilde\Phi^{\dot\alpha\alpha}(\vec\sigma),
L_{\gamma\dot\delta}(\vec\sigma')\}_{P.B.}=2\delta^{\dot\alpha}_{\dot\delta}\widetilde\Phi_{\gamma}{}^\alpha\delta^p(\vec\sigma-\vec\sigma')
\end{array}
\end{equation}
and  conclude that the r.h.s. of (\ref{L}) are zero in the weak sense.

The same conclusion  is satisfied  for both the nonzero P.B. with 
$\widetilde S$ generator densities
%as it follows  from
\begin{equation}\label{L'}
\begin{array}{rl}
\{\widetilde\Phi^{\alpha\beta}(\vec\sigma),
{\widetilde S}_{\gamma}(\vec\sigma')\}_{P.B.}=&
\frac12[\delta^{\alpha}_{\gamma}(\widetilde\Psi^{\beta}+8i\widetilde\Phi^{\beta\delta}\theta_\delta+4i\widetilde\Phi^{\dot\delta\beta}\bar\theta_{\dot\delta})\\[0.2cm]
+&
\delta^{\beta}_{\gamma}(\widetilde\Psi^{\alpha}+8i\widetilde\Phi^{\alpha\delta}\theta_{\delta}+4i\widetilde\Phi^{\dot\delta\alpha}\bar\theta_{\dot\delta})]\delta^p(\vec\sigma-\vec\sigma');\\[0.2cm]
\{\widetilde\Phi^{\dot\alpha\alpha}(\vec\sigma),
S_{\gamma}(\vec\sigma')\}_{P.B.}=&
\delta^{\alpha}_{\gamma}(\bar{\widetilde\Psi}{}^{\dot\alpha}+8i\bar{\widetilde\Phi}{}^{\dot\alpha\dot\delta}\bar\theta_{\dot\delta}+4i\widetilde\Phi^{\dot\alpha\delta}\theta_{\delta})\delta^p(\vec\sigma-\vec\sigma')
\end{array}
\end{equation}
and  for the generalized conformal boost generator densities $\widetilde K$
\begin{equation}\label{K}
\begin{array}{c}
\{\widetilde\Phi^{\alpha\beta}(\vec\sigma),
{\widetilde
K}_{\gamma\delta}(\vec\sigma')\}_{P.B.}=\frac12\left(\delta^{\alpha}_{\gamma}[-2(z_\delta{}^\lambda+2i\theta_\delta\theta^\lambda)\widetilde\Phi_{\lambda}{}^{\beta}+(x_{\delta\dot\lambda}+2i\theta_\delta\bar\theta_{\dot\lambda})\widetilde\Phi^{\dot\lambda\beta}+\theta_\delta\widetilde\Psi^\beta]\right.\\[0.2cm]
\left.+
\delta^{\alpha}_{\delta}[-2(z_\gamma{}^\lambda+2i\theta_\gamma\theta^\lambda)\widetilde\Phi_{\lambda}{}^{\beta}+(x_{\gamma\dot\lambda}+2i\theta_\gamma\bar\theta_{\dot\lambda})\widetilde\Phi^{\dot\lambda\beta}+\theta_\gamma\widetilde\Psi^\beta]+(\alpha\leftrightarrow\beta)\right)\delta^p(\vec\sigma-\vec\sigma'),\\[0.2cm]
\{\widetilde\Phi^{\alpha\beta}(\vec\sigma),
K_{\gamma\dot\gamma}(\vec\sigma')\}_{P.B.}=\frac12\left(\delta^{\alpha}_{\gamma}[2(x_{\lambda\dot\gamma}+2i\bar\theta_{\dot\gamma}\theta_\lambda)\widetilde\Phi^{\beta\lambda}-(\bar
z_{\dot\gamma}{}^{\dot\lambda}+2i\bar\theta_{\dot\gamma}\bar\theta^{\dot\lambda})\widetilde\Phi^{\beta}{}_{\dot\lambda}+\bar\theta_{\dot\gamma}\widetilde\Psi^\beta]\right.\\[0.2cm]
\left.+(\alpha\leftrightarrow\beta)\right)\delta^p(\vec\sigma-\vec\sigma');\\[0.2cm]
\{\widetilde\Phi^{\dot\alpha\alpha}(\vec\sigma),
K_{\gamma\delta}(\vec\sigma')\}_{P.B.}=\left(\delta^{\alpha}_{\gamma}[-(z_\delta{}^\lambda+2i\theta_\delta\theta^\lambda)\widetilde\Phi_{\lambda}{}^{\dot\alpha}+2(x_{\delta\dot\lambda}+2i\theta_\delta\bar\theta_{\dot\lambda})\widetilde\Phi^{\dot\lambda\dot\alpha}+\theta_\delta\bar{\widetilde\Psi}{}^{\dot\alpha}]\right.\\[0.2cm]
\left.+(\gamma\leftrightarrow\delta)\right)\delta^p(\vec\sigma-\vec\sigma'),\\[0.2cm]
\{\widetilde\Phi^{\dot\alpha\alpha}(\vec\sigma),
K_{\gamma\dot\gamma}(\vec\sigma')\}_{P.B.}=\left(\delta^{\alpha}_{\gamma}[(x_{\lambda\dot\gamma}+2i\bar\theta_{\dot\gamma}\theta_\lambda)\widetilde\Phi^{\dot\alpha\lambda}+2(\bar
z_{\dot\gamma\dot\lambda}+2i\bar\theta_{\dot\gamma}\bar\theta_{\dot\lambda})\widetilde\Phi^{\dot\lambda\dot\alpha}+\bar\theta_{\dot\gamma}\bar{\widetilde\Psi}{}^{\dot\alpha}]\right.\\[0.2cm]
\left.+\delta^{\dot\alpha}_{\dot\gamma}[-2(z_\gamma{}^\lambda+2i\theta_\gamma\theta^\lambda)\widetilde\Phi_{\lambda}{}^{\alpha}+(x_{\gamma\dot\lambda}+2i\theta_\gamma\bar\theta_{\dot\lambda})\widetilde\Phi^{\dot\lambda\alpha}+\theta_\gamma\widetilde\Psi^\alpha]\right)\delta^p(\vec\sigma-\vec\sigma').
\end{array}
\end{equation}

Secondly, the $\widetilde\Psi$-constraints have the following nonzero P.B's. 
with the
$OSp(1|8)$ generator densities
\begin{equation}
\begin{array}{c}
\{\widetilde\Psi_\alpha(\vec\sigma),{\widetilde
L}_{\beta\gamma}(\vec\sigma')\}_{P.B.}=\varepsilon_{\alpha\gamma}\widetilde\Psi_\beta\delta^p(\vec\sigma-\vec\sigma'),\
\{\widetilde\Psi_\alpha(\vec\sigma),{\widetilde
L}_{\dot\beta\gamma}(\vec\sigma')\}_{P.B.}=\varepsilon_{\alpha\gamma}\bar{\widetilde\Psi}_{\dot\beta}\delta^p(\vec\sigma-\vec\sigma');\\[0.2cm]
\{\widetilde\Psi_\alpha(\vec\sigma),\widetilde
S_{\gamma}(\vec\sigma')\}_{P.B.}=4i\varepsilon_{\alpha\gamma}(\widetilde\Psi^\beta\theta_\beta+\bar{\widetilde\Psi}{}^{\dot\beta}\bar\theta_{\dot\beta})\delta^p(\vec\sigma-\vec\sigma');\\[0.2cm]
\{\widetilde\Psi_\alpha(\vec\sigma),
\widetilde
K_{\gamma\delta}(\vec\sigma')\}_{P.B.}=\left(\varepsilon_{\alpha\delta}[(z_{\gamma\beta}+2i\theta_\gamma\theta_\beta)\widetilde\Psi^\beta+(x_{\gamma\dot\beta}+2i\theta_\gamma\bar\theta_{\dot\beta})\bar{\widetilde\Psi}{}^{\dot\beta}]+(\gamma\leftrightarrow\delta)\right)\delta^p(\vec\sigma-\vec\sigma'),\\[0.2cm]
\{\widetilde\Psi_\alpha(\vec\sigma),
\widetilde
K_{\gamma\dot\gamma}(\vec\sigma')\}_{P.B.}=\varepsilon_{\alpha\gamma}[(x_{\beta\dot\gamma}+2i\bar\theta_{\dot\gamma}\theta_\beta)\widetilde\Psi^\beta+(\bar
z_{\dot\gamma\dot\beta}+
2i\bar\theta_{\dot\gamma}\bar\theta_{\dot\beta})
\bar{\widetilde\Psi}{}^{\dot\beta}]\delta^p(\vec\sigma-\vec\sigma').
\end{array}
\end{equation}
The above expressions should be complemented by their complex conjugate.

The Poisson brackets of the $OSp(1|8)$ generator densities denoted 
collectively
by $ \widetilde G\!\!=\!\!(Q_\alpha, \bar Q_{\dot\alpha}, {\widetilde
S}_\alpha, \bar{\widetilde S}_{\dot\alpha},
\pi_{\alpha\beta}, \bar\pi_{\dot\alpha\dot\beta}, P_{\alpha\dot\beta},
{\widetilde L}_{\alpha\beta}, \bar{\widetilde L}_{\dot\alpha\dot\beta},
{\widetilde L}_{\alpha\dot\beta},
{\widetilde L}_{\dot\beta\alpha}, {\widetilde K}_{\alpha\beta}, \bar
{\widetilde K}_{\dot\alpha\dot\beta},
{\widetilde K}_{\alpha\dot\beta})$ with the
$\vec\sigma$-re\-pa\-ra\-me\-trization generators $\widetilde L_M$
are equal to
\begin{equation}
\{\widetilde L_M(\vec\sigma),
G(\vec\sigma')\}_{P.B.}=-G(\vec\sigma)\partial_M\delta^p(\vec\sigma-\vec\sigma').
\end{equation}
The $OSp(1|8)$ generator densities strongly commute with  the
converted first-class
constraints $\widetilde\Delta_W$, $\widetilde P_{u\alpha}$, $\bar{\widetilde
P}_{u\dot\alpha}$, $\widetilde P^{(\rho)}_\tau$ and $P^{(\rho)}_M$.

Thus, we conclude that the $OSp(1|8)$ generators are gauge invariant
quantities on the  classical level and  the next step is to study  this
gauge invariance on the quantum level.

%In the next section we will show that it is not  so and the conformal
%anomaly appears.

\section{Quantum superalgebra of global and gauge symmetries in the extended
phase space. Conformal anomaly?}

To lift the considerations of the previous Section to the quantum
level  one has to consider all the quantities entering expressions
for the first-class constraints (\ref{29})-(\ref{32}), (\ref{36}),
(\ref{38}) and $OSp(1|8)$ generator densities (\ref{47}),
(\ref{50}), (\ref{52}), (\ref{53}), (\ref{55}) as operators and to
choose certain ordering prescription for them. The next step of
the quantum theory consistency check is to verify the validity of
$OSp(1|8)$ superalgebra (anti)commutation relations and the gauge
invariant character of generator densities. For definiteness let  
 start below  considering $\hat{\cal Q}\hat{\cal P}$-ordering. So that
constraints and generator densities are the sums of monomials of
the form $\Pi\hat{\cal Q}\Pi\hat{\cal P}=\hat{\cal
Q}^{n_1}_1\hat{\cal Q}^{n_2}_2\ldots\hat{\cal Q}^{n_k}_k\hat{\cal
P}^{m_1}_1\hat{\cal P}^{m_2}_2\ldots\hat{\cal P}^{m_l}_l$ and the
chosen ordering will be preserved in course of calculation of
(anti)commutators if it is preserved in "elementary"
(anti)commutators $[\Pi\hat{\cal Q}, \Pi\hat{\cal P}\}$,
$[\Pi\hat{\cal P}, \Pi\hat{\cal Q}\}$. It was argued in
\cite{BZ} that the sufficient condition for
ordering preservation is the absence for any $\hat{\cal
Q}$-monomial of the corresponding $\hat{\cal P}$-monomial(s)
containing more than one momentum variable conjugate to that of
$\hat{\cal Q}$-monomial. As is readily seen from the expressions
for the constraints (\ref{29})-(\ref{32}), (\ref{36}), (\ref{38})
and  $OSp(1|8)$ generator densities (\ref{47}), (\ref{50}),
(\ref{52}), (\ref{53}), (\ref{55}) there are present only
$\hat{\cal P}$ monomials of  the first power satisfying
automatically the above criterion except for $\hat{\tilde{\tilde
P}}_{u\gamma}\hat{\tilde{\tilde P}}_{u\lambda}$,
$\hat{\tilde{\tilde P}}_{u\gamma}\hat{\bar{\tilde{\tilde
P}}}_{u\dot\gamma}$ monomials entering the generalized conformal
transformations generator densities $\hat{\widetilde
K_{\gamma\lambda}}(\tau,\vec\sigma)$, $\hat{\widetilde
K_{\gamma\dot\gamma}}(\tau,\vec\sigma)$ (\ref{52}), (\ref{53}).
Their conjugate $\hat{\cal Q}$ monomials $\hat{\tilde
u}_\alpha\hat{\tilde u}_\beta$, $\hat{\tilde
u}_\alpha\hat{\bar{\tilde u}}_{\dot\beta}$ enter
$\hat{\widetilde\Phi}$-constraints (\ref{31}) and direct calculation
reveals anomalous contributions (with $\hbar=c=1$) to the classical 
expressions (\ref{K})
\begin{equation}\label{66}
\begin{array}{c}
[\hat{\widetilde\Phi}{}^{\alpha\beta}(\vec\sigma),
\hat{\widetilde
K}_{\gamma\lambda}(\vec\sigma')]|_{\mbox{anomal. 
contr.}}=-\frac{1}{2}(\delta^\alpha_\gamma\delta^\beta_\lambda+\delta^\alpha_\lambda\delta^\beta_\gamma)\delta^p_\epsilon(\vec\sigma-\vec\sigma')\delta^p_\epsilon(0),\\[0.2cm]

[\hat{\widetilde\Phi}{}^{\dot\alpha\alpha}(\vec\sigma),\hat{\widetilde
K}_{\gamma\dot\gamma}(\vec\sigma')]
|_{\mbox{anomal. 
contr.}}=-\delta^\alpha_\gamma\delta^{\dot\alpha}_{\dot\gamma}\delta^p_\epsilon(\vec\sigma-\vec\sigma')\delta^p_\epsilon(0)
\end{array}
\end{equation}
and their c. c. relations, where the P.B's. were changed by  the commutators
\begin{equation}\label{67}
[\hat{\tilde{\tilde P}}_{u\alpha}(\vec\sigma),\hat{\tilde
u}{}^\beta(\vec\sigma')]=i\delta^\beta_\alpha\delta^p_\epsilon(\vec\sigma-\vec\sigma'),\
[\hat{\bar{\tilde{\tilde P}}}_{u\dot\alpha}(\vec\sigma),\hat{\bar{\tilde
u}}{}^{\dot\beta}(\vec\sigma')]=i\delta^{\dot\beta}_{\dot\alpha}\delta^p_\epsilon(\vec\sigma-\vec\sigma')
\end{equation}
including the regularized delta function $\delta^p_\epsilon(\vec\sigma-\vec\sigma') $ considered in  \cite{ILST}.
This means that the quantum conformal boosts are not gauge invariant
operators and the generalized conformal symmetry is broken. 
However, this  breaking could be a consequence of the $\hat{\cal Q}\hat{\cal P}$-ordering prescription used for the  transition to quantum theory\footnote{It was noted by  D. Sorokin with the reference on the  paper \cite{Vasiliev}, where a modified realization of the conformal generators was considered for the quantum  superparticle without breaking of the conformal supersymmetry.}. In fact, the anomalous terms may be hidden if we consider the hermitian expressions for the generator densities. These expressions differ from the above considered  $\hat{\cal Q}\hat{\cal P}$-ordered expressions
(\ref{47}),(\ref{50}), (\ref{52}), (\ref{53}), (\ref{55})  by the addition
of singular terms proportional to 
$\delta^p_\epsilon(0)$
\begin{equation}\label{68}
\begin{array}{c}
\hat{\widetilde{\mathcal S}}_{\gamma}=\hat{\widetilde S}_{\gamma}+ 
6\theta_\gamma\delta^p_\epsilon(0),\   
\hat{\bar{\widetilde{\mathcal S}}}_{\dot\gamma}=
\hat{\bar{\widetilde S}}_{\dot\gamma} +6\bar{\theta}_{\dot\gamma}\delta^p_\epsilon(0),\\
\hat{\widetilde{\mathcal K}}_{\gamma\lambda}=\hat 
{\widetilde K}_{\gamma\lambda}-3iz_{\gamma\lambda}\delta^p_\epsilon(0),\ 
\hat{\bar{\widetilde{\mathcal K}}}_{\dot\gamma\dot\lambda}=\hat{\bar {\widetilde K}}_{\dot\gamma\dot\lambda}-3i\bar z_{\dot\gamma\dot\lambda}\delta^p_\epsilon(0),\ 
\hat{\widetilde{\mathcal K}}_{\gamma\dot\gamma}=\hat {\widetilde K}_{\gamma\dot\gamma}-3ix_{\gamma\dot\gamma}\delta^p_\epsilon(0),\\
\hat{\widetilde{\mathcal L}}_{\alpha\beta}=\hat {\widetilde L}_{\alpha\beta}-\frac{3i}{2}\varepsilon_{\alpha\beta}\delta^p_\epsilon(0),\ 
\hat{\bar{\widetilde{\mathcal L}}}_{\dot\alpha\dot\beta}=\hat{\bar{\widetilde L}}_{\dot\alpha\dot\beta}-\frac{3i}{2}\bar{\varepsilon}_{\dot\alpha\dot\beta}\delta^p_\epsilon(0).
\end{array}
\end{equation}
These singular terms appear as a  result of the $\hat{\cal Q}\hat{\cal P}$-ordering in the non-ordered hermitian expressions for  the generators and using 
 the  relations similar to 
\begin{equation}\label{69}
\frac12(z_{\lambda\delta}(\vec\sigma)\pi^{\delta\varepsilon}(\vec\sigma)+\pi^{\delta\varepsilon}(\vec\sigma)z_{\lambda\delta}(\vec\sigma))=z_{\lambda\delta}(\vec\sigma)\pi^{\delta\varepsilon}(\vec\sigma)-\frac{3i}{4}\delta_\lambda^\varepsilon\delta^p_\epsilon(0)
\end{equation}
for all products of the brane coordinates and momenta. Then  we find that the substitution of $\hat{\widetilde{\mathcal K}}$ (\ref{68}) for $\hat{\widetilde K}$  in the commutators  (\ref{66}) will 
change only the coefficients there 
%inuse the modified expressions (\ref{68}) will change the commutators (\ref{66%}) to the form 
\begin{equation}\label{70}
\begin{array}{c}
[\hat{\widetilde\Phi}{}^{\alpha\beta}(\vec\sigma),
\hat{\widetilde{\mathcal 
K}}_{\gamma\lambda}(\vec\sigma')]|_{\mbox{anomal. 
contr.}}=-(\frac{1}{2}+\frac{3}{2})(\delta^\alpha_\gamma\delta^\beta_\lambda+\delta^\alpha_\lambda\delta^\beta_\gamma)\delta^p_\epsilon(\vec\sigma-\vec\sigma')\delta^p_\epsilon(0),
\\[0.2cm]
[\hat{\widetilde\Phi}{}^{\dot\alpha\alpha}(\vec\sigma),\hat{\widetilde{\mathcal K}}_{\gamma\dot\gamma}(\vec\sigma')]
|_{\mbox{anomal. 
contr.}}=-(1+3)\delta^\alpha_\gamma\delta^{\dot\alpha}_{\dot\gamma}\delta^p_\epsilon(\vec\sigma-\vec\sigma')\delta^p_\epsilon(0)
\end{array}
\end{equation}
and in their c. c. commutators. To derive  this result the changes $\delta^p(\vec\sigma-\vec\sigma') \rightarrow \delta^p_\epsilon(\vec\sigma-\vec\sigma') $ and i$\{\cal P,\cal Q \}_{P.B.} \rightarrow$ [ $\hat{\cal P}  ,\hat{\cal Q}$ ]
 in the canonical P.B's. (\ref{12}) were taken into account. 

Then we revealed  that the singular terms in the r.h.s. of (\ref{70}) are precisely the terms which are needed for the hermiticity restoration of the  $\hat{\cal Q}\hat{\cal P}$-ordered  non-anomalous operators in the r.h.s. (\ref{K}).
We found  that the same  effect is  valid for other (anti)commutators of the algebra.  Thus, we have  shown that the 
considered singular terms in (\ref{66}) and (\ref{70}) may be hidden in the expressions for the disordered hermitian generators. If we have started  from 
this  hermitian representation we could not observe the singular terms. 

So, it seems that namely these non-ordered definitions might be accepted as correct quantum definitions of the conformal hermitian generators. 
However, it is not end of the story, because of the necessity to define the relevant vacuum state and to construct both the proper physical space of states associated with the considered BPS branes and nilpotent BRST operator. 
This  problem is open for the string or  brane in contrast to  the superparticle \cite{Vasiliev}, because  there the $\hat{\cal Q}\hat{\cal P}$-ordered 
   hermitian   superconformal generators contain only  finite central terms instead of the  singular terms under question. It is  clear, because $p$-brane is  a field object with infinite number of degrees of freedom. 
Similar  problem was studied in the series of papers \cite{ILST},\cite{GLSSU} devoted to the light-cone quantization of tensionless string.  The starting point there  was  consideration of hermitian expressions for the generators and constraints which have  further been  transformed to the $ \hat{\cal Q}\hat{\cal P}$-order. This ordering was chosen as physical one consistent with the definition of the vacuum state as a state annihilated by the momentum operator. 
The  definition \cite{GLSSU} of the vacuum state was physically justified by the analysis of the tensionless limit of the vacuum conditions accepted in the tensile string theory. As  a result, the conformal symmetry breaking was revealed there for the space-time  dimension $D>2$.  

\section{Conclusion}

In the present report we studied  the classical and quantum symmetries  of
the recently proposed tensionless super $p$-brane  model \cite{ZU} in $D=4$ 
$N=1$ superspace extended by TCC  coordinates $z_{mn}$.
The primary and secondary constraints of the model form a  
 mixed set of the first- and second-class constraints and the Dirac bracket 
realization of the global and gauge symmetries results in rather  nonlinear 
expressions obtained in  \cite{UZ03}.
One of the possible methods to get rid of the second-class constraints,  and 
respectively to avoid  the  nonlinearities,  is
their conversion  to the first-class constraints using 
additional conversion variables \cite{Fad}-\cite{EM} and preserving the number of  the primary  physical degrees of freedom.
We applied this method here to  the primary and secondary  constraints of  
the  model \cite{ZU} and transformed them into a set of the converted first-class constraints.  As a result, we found the classical realization of the $OSp(1|8)$ generators in extended phase space and established their invariance under the  world-volume symmetry transformations.

For the  transition to quantum theory we studied the problem of ordering of the generalized $ \hat{\cal Q}$ and $\hat{\cal P}$  operators of the brane. 
We found that the  hermitian expressions for   $OSp(1|8)$ generators without $\hat{\cal Q}\hat{\cal P}$-ordering of the operators forming them may be considered as a quantum generalization of the  corresponding  classical generators.
 Together with the generators of the gauge  world-volume symmetry they  form closed  quantum algebra of local and global brane  symmetries free of anomalous terms. 

However, the problem appears how to construct the relevant vacuum state and  proper physical space for the $\hat{\cal Q}\hat{\cal P}$-desordered quantum generators and constraints.
 Transition to the $\hat{\cal Q}\hat{\cal P}$-ordering in the hermitian expressions for the $OSp(1|8)$ quantum generators  is probably able to weaken  the mentioned problem, but then we need to give a proper physical interpretation of the above mentioned singular terms in the generators and  commutators.
It is possible that the analyzed  quantum deviations  could imply  explicit breakdown of the classical superconformal symmetry  $OSp(1|8)$ after quantization if the  $\hat{\cal Q}\hat{\cal P}$-ordering might be approved as physical one.
 That  possibility has been realized in \cite{GLSSU} 
for tensionless strings, having no oscillator degrees of freedom, where the  $\hat{\cal Q}\hat{\cal P}$-ordering was justified to be  physical. It was shown there that this ordering is consistent with the existence of nilpotent BRST operator and the absence of divergent anomalous  terms  only in the space-time with dimension $D=2$.
Note that, as it follows from (\ref{68}), the contributions of the spinor, vector and tensor  degrees of freedom to the  brane anomaly were obtained namely 
 for the  $D=4$ case which is in the black list of \cite{GLSSU}. Let us also remind  that the conformal invariance  for massless particle in $D>2$ survives the transition to the quantum case, as it was also subscribed  in \cite{GLSSU}.  This conclusion agrees with the above mentioned result of \cite{Vasiliev}, where superconformal generators of superparticle include only  finite terms  resulting from the  $\hat{\cal Q}\hat{\cal P}$-ordering. For the brane case, however the  similar terms  are divergent, as it follows from (\ref{68}) and (\ref{69})
 if  the regularization  in the delta functions is removed, i.e. the limit  $\epsilon \rightarrow 0 $ is implemented.
Thus, to prove that  $OSp(1|8)$ conformal supersymmetry survives the quantization with  the considered disordered hermitian $OSp(1|8)$ generators and world-volume gauge constraints one has to construct proper  vacuum state and physical space of states permitting nilpotent BRST charge.
 
As a result, the question whether the  global $OSp(1|8)$ supersymmetry is proper quantum  symmetry of the exotic BPS branes needs in further investigation.
 Finally note that  it is very interesting to apply our analysis to exotic $OSp(1|64)$ invariant  superbranes of $M$-theory in  $N=1 \, D=11$ superspace enlarged by TCC coordinates.

\section{Acknowledgements}

The authors are grateful to the Organizers of SQS Workshop for the
warm hospitality in Dubna and I.~Bengtsson, A. Laudal, M. Movshev, A. Pashnev,  D. Sorokin  and B. Sundborg for useful remarks and fruitful discussions.
A.Z. thanks Fysikum at the Stockholm University and the Mittag-Leffler Institute for kind hospitality.  The work was 
partially supported by the grant of the Royal Swedish Academy of Sciences 
and by the SFFR of  Ukraine under Project 02.07.276.

\end{document}